# A Social Recommender System based on Bhattacharyya Coefficient

Mohammad Reza Zarei
Department of
Computer Science,
Engineering and IT
Shiraz University
Shiraz, Iran

Mohammad Reza Moosavi
Department of
Computer Science,
Engineering and IT
Shiraz University
Shiraz, Iran

*Abstract* — **Recommender systems play a significant role in providing the appropriate data for each user among a huge amount of information. One of the important roles of a recommender system is to predict the preference of each user to some specific data. Some of these systems concentrate on user-item networks that each user rates some items. The main step for item recommendation is to predict the rate of unrated items. Each recommender system utilizes different criteria such as the similarity between users or social relations in the process of rate prediction. As social connections of each user affect his behaviors, it can be a valuable source to use in rate prediction.**

**In this paper, we will provide a new social recommender system which uses Bhattacharyya coefficient in similarity computing to be able to evaluate similarity in sparse data and between users without co-rated items as well as integrating social ties into the rating prediction process.**

## 1. Introduction

Nowadays, recommender systems are so purposive and widely used in different areas from commercial means as offering products to social friend recommendation. Their usages are significantly increasing day by day. One of the applications of recommender systems is to suggest new items or products to a user, based on different characteristics. For this purpose, the primary phase is to predict the user's rate for each item that is still unrated. Next, the system could suggest highly rated items to the user.

To predict the item rate, similarity-based recommender systems need to calculate the similarity between a query (i.e., a user that want to do prediction for) and all user who has already rated the item.

Various similarity measures are used in recommender systems. Some popular ones such as cosine similarity [1] and Pearson correlation [2] that use co-rated items for calculating similarity between users, suffer from few or none common items [3] that may cause the recommender system unable to calculate similarity between users that have rated acceptable number of items but without enough common ones. Therefore, they are impotent in cold-start and sparse data state.

In recent years, researchers had become interested in similarity measurements that use more than co-rated items for similarity calculation. This will end to much more capability in similarity estimation in sparse data.

Also utilizing social relations among users has attracted wide attention in recent works not only to prevent cold start problem from deteriorating the performance of recommender systems [8], but also to make the whole recommendation process outperform. Since people's activities and choices could have great impact on each other, the significance of using social ties in recommender systems has arisen in recent years that can be even more effective than first-hand similar users.

Therefore, utilizing the social ties as well as forming the right method for similarity evaluation could lead to a great performance in recommender systems

In this paper, we will introduce a new rating-prediction method using Bhattacharyya coefficient which also utilizes social relations between users. It uses all ratings made by each pair of users to calculate the similarity and also integrates social ties into the process to get the best result.

This work makes the following contributions:
- We propose a similarity calculation method based on Bhattacharyya Coefficient that uses all ratings of each pair of users instead of using just co-rated items.
- We will integrate social ties into our method as well as using direct similarity of user with others.
- We will improve our methodology by substituting the similarity with difference that is a better metric to reflect the contrast between users.

## 2. Background

**Bhattacharyya Coefficient**

Bhattacharyya Coefficient (BC) [7] is a metric to determine the closeness of two probability distributions. It is defined [3,4] as:

$$BC(p_1, p_2) = \sum_{x \in X} \sqrt{p_1(x) p_2(x)} \qquad (1)$$

where $p_1(x)$ and $p_2(x)$ are two density distribution over a discrete domain X.

*BC* similarity between two items *i* and *j* with estimated discrete densities $\hat{p}_i$ and $\hat{p}_j$ can be computed as [5,6]:

$$BC(i,j) = BC(\hat{p}_i, \hat{p}_j) = \sum_{h=1}^{m} \sqrt{(\hat{p}_{ih})(\hat{p}_{jh})} \quad (2)$$

where *m* is the number of bins and $\hat{p}_{ih} = \frac{\#h}{\#i}$, where #i is the total number of users rated the item *i* and #h is the number of users rated item *i* with rating score '*h*'. For instance, if 20 users had rated item *x*, and 5 of them had rated with the score 3, $\hat{p}_{x3}$ will be $\frac{5}{20}$. The maximum value for BC is 1.

Patra et al. in [3] proposed a similarity measure based on Bhattacharyya coefficient that uses all ratings made by a pair of users. He used the Eq. (2) to calculate the global similarity of two items. We will also use this equation in users' similarity evaluation for our rating prediction method in the following section.

## 3. Proposed Method

In this section, we will explain our approach for item-rating prediction. Two methods will be proposed to do this prediction while the second method is the improved version of the first one. In section 4, the results of them will be illustrated and compared.

### 3.1 Method A

Let $I_q$ be the list of items rated by user *q* and $r_{qi}$ be the rate value given by user *q* to item *i*.

The similarity between user *q* and user *x* is calculated as:

$$SIM(q,x) = \frac{\sum_{i \in I_q} \sum_{j \in I_x} BC(i,j) RSP(r_{qi}, r_{xj})}{\sum_{i \in I_q} \sum_{j \in I_x} BC(i,j)} \quad (3)$$

where $RSP(r_{qi}, r_{xi})$ is the rate similarity between $r_{qi}$ and $r_{xi}$ and is calculated as:

$$RSP(r_{qi}, r_{xj}) = 1 - (|r_{qi} - r_{xj}| * m) \quad (4)$$

where *m* is the amount that decreases RSP with each one unit difference between two rates. For instance, if we consider the range of rates from integer values [1,5], the maximum difference between two rates is 4 and the differences unit is 1. As a result, each one unit difference between two rates will reduce RSP by $\frac{1}{4}$, so *m=0.25*. If two rates are the same, RSP will be equal to 1 that is the maximum similarity between two rate values.

To predict the rating value of an item given by a user, direct similarity of the user with each scorer of the item is calculated using Eq. (3). Also the overall similarity of social ties with each scorer will be considered.

#### 3.1.1 Social Ties Integration

To use overall trusted users' opinion about a scorer, the similarity between the set and each scorer is needed to be calculated.

Let $T_q$ be the set of users trusted by user *q*. The similarity between this set and a user *t* is calculated as:

$$TSIM(T_q, t) = \frac{\sum_{x \in T_q} SIM(x,q) SIM(x,t)}{\sum_{x \in T_q} SIM(x,q)} \quad (5)$$

which is the weighted average of similarities between each trusted user and the user *t* (i.e., *SIM(x,t)*). The weight factor of this average is similarity between each trusted user and the trustor (*SIM(x,q)*).

For rate prediction, *TSIM* will be integrated into the direct similarity between the user and scorer. The formulation is provided in the following subsection.

#### 3.1.2 Rate Prediction

Suppose that we want to predict the rating of user *q (query user)* to item *t* ($\hat{r}_{gt}$). For each member of $U_t$ (called user *y*), direct similarity with user q and similarity with $T_q$ will be merged as the average of *SIM(q,y)* and *TSIM($T_q,y$)*.

$$AVG(SIM(q,y), TSIM(T_q,y)) = \frac{SIM(q,y) + TSIM(T_q,y)}{2} \quad (6)$$

In cases that calculating one of these similarities is impossible, the other will be used. Eventually, the predicted rate is calculated as:

$$approximate\ \hat{r}_{qt} = \frac{\sum_{y \in U_t}(AVG(SIM(q,y), TSIM(T_q,y)) \times r_{yt})}{\sum_{y \in U_t} AVG(SIM(q,y), TSIM(T_q,y))} \quad (7)$$

This value will be rounded to nearest valid rate.

#### 3.1.3 Method A Imperfection

As the *RSP* factor used in similarity evaluation equation (Eq. (3)) uses absolute value to measure the difference of rate values, it does not reflect the real diversity. For instance, *RSP* will treat two rate values 3 and 2 the same as 2 and 3 and it will output 0.75 as the rate values' similarity. As we want to tune each rating for the intended item in rate prediction based on the similarity between scorer's rates and user's rates, it is important for similarity evaluation process to specify the positivity or negativity of difference in rating values as well as the absolute value. So, we propose method B which utilizes the differences between users' rates in rating prediction process.

### 3.2 Method B

In method A, we calculate the similarity of two users based on the *BC* similarity between the rated items and the similarity between the rate values (*RSP*). In this method we will calculate the weighted average of differences between each rate values of user *g* and user *x* and use the real difference of values instead of using *RSP*. This difference will reflect the diversity of rates better than the similarity that does not highlight the positivity or negativity.

```
 1:  function CALCULATE_RATE (q, i, U_i, T, R)              Returns rate
     ▷ Input: q: the user we want to predict for (query user), i: the intended item, U_i: list of users who have rated item i
       with their corresponding rate, T: set of users trusted by query user (q), R: list of all rates of each user
     ▷ Output: rate: the predicted rate
 2:  I_q = R[q]                                              ▷ list of all items rated by q and his rates
 3:  for each scorer x in U_i do
 4:         I_x = R[x]
 5:         diff_q = CALCULATE_USERS_DIFFERENCE (q, x, I_q, I_x)                                      Eq.(8)
 6:         diff_trustees = CALCULATE_TRUSTEES_DIFFERENCE_WITH_SCORER (T, q, x, S_T, I_q, I_x)        Eq.(9)
 7:         diff = 0
 8:         if diff_trustees != 0 and diff_q != 0 then     diff = (diff_trustees + diff_q)/2
 9:         else if diff_q != 0 then                        diff = diff_q
10:         else if diff_trustees != 0 then                 diff = diff_trustees
11:         rate_numerator += (r+diff)                      ▷ r is the rate of scorer x to item i
12:         rate_denominator += 1
13: rate = rate_numerator / rate_denominator
14: round rate to the nearest integer
15: if rate > 5 then       rate = 5
16: else if rate < 1 then  rate = 1
17: return rate
```

**Figure 1: Method B algorithm**

The algorithm of Method B is presented in Fig. (1). In this algorithm, the difference level of two users $q$ and x is calculated as:

$$DIF(q,x) = \frac{\sum_{i \in I_q} \sum_{j \in I_x} BC(i,j)(r_{qi} - r_{xj})}{\sum_{i \in I_q} \sum_{j \in I_x} BC(i,j)} \quad (8)$$

In this equation, *BC* of each two items is the weight of the values' difference in the average calculation. Generally, Eq. (8) will be used to determine the value which is needed to be added to the scorer's rate to be the predicted value. It calculates difference of query user with each scorer (See line 5 in Fig. (1)). We also use this equation to integrate trusted users' difference with scorer which is explained in next subsection.

**3.2.1 Social Ties Integration**

In addition to use the difference between each scorer and the intended user, we utilize the difference of social ties and the scorers.

The difference between trusted users by user $q$ ($T_q$) and each scorer (called $t$) is calculated as (line 6 of Fig. (1)):

$$TDIF(T_q, t) = \frac{\sum_{x \in T_q} SIM(x,q) DIF(x,t)}{\sum_{x \in T_q} SIM(x,q)} \quad (9)$$

As it is obvious, *TDIF* is the weighted average of differences between each trusted user and the scorer (using *DIFF*), while the weight factor is the similarity between trusted user and who trusts which is calculated by Eq (3).

**3.2.2 Rate Prediction**

To predict the rate of item $t$ by user $q$, for each member of $U_t$ (called $y$), $AVG(DIF(q,y), TDIF(T_q,y))$ is calculated and this value is added to $r_{yt}$. If AVG value is unavailable to calculare, it will be replaced by 0. In another words, we are tuning each rate with $AVG(DIF(q,y), TDIF(T_q,y))$ and if this value is not available, the own rate is used (See lines 7-10).

Note that each one of the differences could be negative or positive. Then, these generated rates will be averaged to form $\hat{r}_{qt}$. The formulation is:

$$approximate \; \hat{r}_{qt} = \frac{\sum_{y \in U_t} r_{yt} + AVG(DIF(q,y), TDIF(T_q,y))}{|U_t|} \quad (10)$$

where $|U_t|$ is the number of members of $U_t$. The value obtained from Eq. (10) will be rounded to nearest valid rate. If it is greater than the maximum possible rate or less than the minimum possible one, it should be reduced or added to reach the nearest valid rate (Line 15 and 16 in Fig. (1)). In cases that calculating one of *DIF* or *TDIF* is impossible, the other will be used instead of using the average and if both of them are unavailable, AVG value in Eq. (10) will be replace by 0.

## 4. Experimental Evaluation

### 4.1. Datasets

We used Epinions[1] real dataset which consists of user-item ratings in range of integer values [1,5] and user-user trust relationships. This dataset is more than 99% sparse. The statistics of the dataset is given in table 1.

| Dataset | #users | #items | #rating | #trust-relation |
|---|---|---|---|---|
| Epinions | 49290 | 139738 | 664824 | 487181 |

**Table 1: Description of datasets**

### 4.2. Evaluation Metrics

To evaluate the performance of our method, we use two metrics, Mean Absolute Error (MAE) and Root Mean Square Error (RMSE). These metrics are defined as follows:

$$\text{MAE} = \frac{\sum |r_{it} - \hat{r}_{it}|}{N}$$

$$\text{RMSE} = \sqrt{\frac{\sum (r_{it} - \hat{r}_{it})^2}{N}}$$

where $r_{it}$ is the rate given by user i to item t and $\hat{r}_{it}$ is the predicted rating of user i for item t. *N* is the number of all predicted items.

### 4.3. Experimental Results

We randomly split user-item ratings into five equal sub-datasets. Each time, we use one of them for testing and the remaining for training. The current result of our method is obtained from two out of five parts being training in turn.

To evaluate the impact of utilizing social ties for each method A and B, we implemented two versions: a non-social version and another one with integrating social ties.

1. **Social version of method A:** This version used trust relations as in Eq. (7).

2. **Non-social version of method A**: In this version, instead of using the average of *SIM* and *TSIM* in rate prediction (Eq. (7)), only *SIM* was used. In another words, we eliminated the impact of social ties in rating prediction by substituting $AVG(SIM(q,y),TSIM(T_q,y))$ with $SIM(q,y)$ in Eq. (7).

3. **Social version of method B:** In this version, we integrated social ties in rating prediction as in Eq. (10).

4. **Non-social version of method B:** This version of method B, changed Eq. (10) to omit the efficacy of social ties by replacing $AVG(DIF(q,y),TDIF(T_q,y))$ with $DIF(q,y)$.

Also we will compare our implemented methods with some state-of-the-art social recommendation approaches:

- TrustMF: A social method capable of integrating sparse rating data given by users and sparse social trust network among these same users proposed by Yang et.al. in [9].
- TrustPMF: This method provides a probabilistic interpretation to TrustMF model [9].
- SoRec: A matrix factorization framework with social regularization [10].
- RSTE: A probabilistic factor analysis framework which naturally fuses the users' tastes and their trusted friends' favors together [11].
- TCF: A trust-aware collaborating filtering-based recommender system [13].
- BIBR: A Bayesian-inference based recommendation system that leverages the embedded social structure [12].

Table 2 illustrates the performance of our methods, TrustMF, TrustPMF, SoRec, RSTE, TCF and BIBR.

| Methods | Coverage percentage | MAE | RMSE |
|---|---|---|---|
| Method A non-social version | 85.34% | 0.7959 | 1.1280 |
| Method A social version | 85.34% | 0.7938 | 1.1299 |
| Method B non-social version | 86.7% | 0.7889 | 1.1328 |
| Method B social version | 86.7% | **0.7837** | **1.1188** |
| TrustMF | 100% | 1.0142 | 1.2725 |
| TrustPMF | 100% | 0.9396 | 1.1674 |
| SoRec | 100% | 1.0401 | 1.2757 |
| RSTE | 100% | 1.0281 | 1.3290 |
| TCF | 77% | 0.805 | NA |
| BIBR | 74.37% | 0.791 | NA |

---

[1] http://www.trustlet.org/downloaded_epinions.html

**Table 2: MAE, RMS and coverage comparison of our methods with some state-of-the-art methods.**

All versions of our method have a test score greater than 85%, while they could be extended with techniques such as using average ratings of users or items for the remaining small unpredicted part to complete the predictions. As we wanted to investigate our algorithms, we did not integrate these strategies into our methods.

According to both MAE and RMSE metrics, the social version of method B outperforms the other implemented versions as we expected due to using difference value instead of absolute value. The social version of method A wins over the non-social version based on MAE, while it is opposite based on RMSE unexpectedly.

We also compared the proposed approach with some state of the art methods (namely TrustMF, TrustPMF, SoRec, RSTE, TCF and BIBR). According to both MAE and RMSE metrics, our method outperforms all of these methods.

The advantage of model-based methods such as TrustMF and TrustPMF is full prediction coverage, while they are impotent in sparse datasets in comparison with similarity based methods. In contrast, similarity-based methods such as TCF excel in sparse datasets while they own a lower coverage percentage. According to the results presented in Table. 2, our method not only improves the accuracy of similarity-based methods in sparse data, but also increases the coverage percentage to meet both high accuracy and coverage.